\documentclass{appolb}
\usepackage{epsfig}
\begin{document}
% \eqsec  % uncomment this line to get equations numbered by (sec.num)
\title{ High density behaviour of nuclear symmetry energy }
% you can use '\\' to break lines}
{
\author{  D.N. Basu \thanks{E-mail: dnb@veccal.ernet.in} and Tapan Mukhopadhyay \thanks{E-mail: tkm@veccal.ernet.in} 
\address{Variable Energy Cyclotron Centre, 1/AF Bidhan Nagar, Kolkata 700 064, India\\}
}
\vskip 0.2cm
\maketitle
\begin{abstract}

      Role of the isospin asymmetry in nuclei and neutron stars, with an emphasis on the density dependence of the nuclear symmetry energy, is discussed. The symmetry energy is obtained using the isoscalar as well as isovector components of the density dependent M3Y effective interaction. The constants of density dependence of the effective  interaction are obtained by reproducing the saturation energy per nucleon and the saturation density of spin and isospin symmetric cold infinite nuclear matter. Implications for the density dependence of the symmetry energy in case of a neutron star are discussed, and also possible constraints on the density dependence obtained from finite nuclei are compared. 

\noindent 
%Keywords : Symmetry energy; Equation of state; Neutron-rich matter; Neutron stars
 
\end{abstract}
\PACS{ 21.65.+f, 26.60.+c, 97.60.Jd, 21.30.Fe }

%\documentstyle[preprint,aps]{revtex}
%\documentstyle[article,eqsecnum,aps]{revtex}
%\documentstyle[twocolum,prc]{revtex}
%\documentstyle[preprint,eqsecnum,aps]{revtex}
%\begin{document}
%\draft

%\date{\today }

%\noindent
%PACS numbers: 21.10.Dr, 21.60.Ev, 21.65.+f      

%\maketitle
%\eject

\section{Introduction}
\label{section1}

      The symmetry energy is an important quantity in the equation of state of isospin asymmetric nuclear matter. This currently unknown quantity plays a key role to the understanding of the structure of systems as diverse as the neutron rich nuclei and neutron stars. Theoretical studies based on microscopic and many-body calculations and phenomenological approaches predict various forms of the density dependence of the symmetry energy. In general, they can be broadly classified into two different forms. One, where the symmetry energy increases monotonically with increasing density (`stiff' dependence) \cite{Ch05} and the other, where the symmetry energy increases \cite{Ba02} initially up to normal nuclear density or slightly beyond and then decreases at higher densities (`soft dependence'). Determination of the exact form of the density dependence of the symmetry energy is of utmost importance for studying the structure of neutron rich nuclei, and studies relevant to astrophysical problems, such as the structure of neutron stars and the dynamics of supernova collapse \cite{Ba01}. For example, a `stiff' density dependence of the symmetry energy is predicted to lead a large neutron skin thickness compared to a `soft' dependence and can result in rapid cooling of a neutron star, and larger neutron star radius, compared to a soft density dependence.   

      In view of rather large differences between the various calculations of the symmetry energy present even at the subsaturation densities, the question arises whether one can obtain empirical constraints from finite nuclei. Since the degree of isospin diffusion in heavy-ion collisions at intermediate energies is affected by the stiffness of the nuclear symmetry energy, these reactions, therefore, can also provide constraints on the low energy behaviour of the nuclear symmetry energy \cite{Ch05}. However, the high density behaviour remains largely undetermined since the masses and corresponding radii of the neutron stars are not measured whereas they can be obtained theoretically by solving Tolman-Oppenheimer-Volkov equation. At present there exist only indirect indications such as the neutron star cooling process. Recently search for the experimental signatures of the moderately high density behaviour of the nuclear symmetry energy has been proposed \cite{Ba02} theoretically using several sensitive probes suh as the $\pi^-$ to $\pi^+$ ratio, tranverse collective flow and its excitation function as well as the neutron-proton differential flow. In the present work, the nuclear symmetry energy is calculated theoretically using the isoscalar and isovector components of M3Y-Reid-Elliott effective interaction supplemented by a zero range pseudo-potential \cite{Sa79} along with the density dependence and its high density behaviour is explored. 
     
\section{The nuclear symmetry energy}
\label{section2}

      The nuclear symmetry energy (NSE) $E_{sym}(\rho)$ is defined as the energy required per nucleon to change the spin and isospin symmetric nuclear matter (SNM) to the spin symmetric pure neutron matter (PNM). In the present work the NSE is obtained using the density dependent M3Y interaction supplemented by a zero-range pseudo-potential (DDM3Y) as the effective interaction between two interacting nucleons inside nuclear medium. The M3Y interaction was derived by fitting its matrix elements in an oscillator basis to those elements of the G-matrix obtained with the Reid-Elliott soft-core NN interaction. The ranges of the M3Y forces were chosen to ensure a long-range tail of the one-pion exchange potential as well as a short range repulsive part simulating the exchange of heavier mesons \cite{Be77}. The real part of the nuclear interaction potential obtained by folding in the density distribution functions of two interacting nuclei with the DDM3Y effective interaction was shown to provide good descriptions for medium and high energy $\alpha$ and heavy ion elastic scatterings \cite{Sa79,Ko84}. The zero-range pseudo-potential represented the single-nucleon exchange term while the density dependence accounted for the higher order exchange effects and the Pauli blocking effects. The real part of the proton-nucleus interaction potential obtained by folding in the density distribution function of interacting nucleus with the DDM3Y effective interaction is found to provide good descriptions of elastic and inelastic scatterings of high energy protons \cite{Gu05} and proton radioactivity \cite{BCS05}. 

      The central part of the effective interaction between two nucleons 1 and 2 can be written as  \cite{Sa79} 

\begin{equation}
 v_{12}(s) = v_{00}(s) + v_{01}(s) \tau_1.\tau_2 + v_{10}(s) \sigma_1.\sigma_2 
+ v_{11}(s) \sigma_1.\sigma_2~\tau_1.\tau_2,
\label{seqn1}
\end{equation}   
\noindent 
where $\tau_1,\tau_2$ are the isospins and $\sigma_1,\sigma_2$ are the spins of nucleons 1,2 and $s$ is the distance between them. In case of SNM, only the first term, the isoscalar term, contributes whereas for the isospin asymmetric-spin symmetric nuclear matter only first two terms, the isoscalar and the isovector terms, contribute and for the spin-isospin asymmetric nuclear matter all the four terms of Eq.(1) contribute.  Considering only the isospin asymmetric-spin symmetric nuclear matter, the neutron-neutron, proton-proton, neutron-proton and proton-neutron interactions, viz. $v_{nn}, v_{pp}, v_{np}$ and $v_{pn}$ respectively, can be given by the following : 

\begin{equation}
 v_{nn} = v_{pp} =v_{00} + v_{01},~~~~v_{np} = v_{pn} = v_{00} - v_{01}.
\label{seqn2}
\end{equation}   
\noindent 
The general expression for the density dependent effective NN interaction potential is written as \cite{BCS05} 

\begin{equation}
 v_{00}(s,\rho, \epsilon) = t_{00}^{M3Y}(s, \epsilon) g(\rho, \epsilon),~~~~v_{01}(s,\rho, \epsilon) = t_{01}^{M3Y}(s, \epsilon) g(\rho, \epsilon),
\label{seqn3}
\end{equation}   
\noindent
where $\epsilon=E/A$ is the energy per nucleon and the isoscalar $t_{00}^{M3Y}$ and the isovector $t_{01}^{M3Y}$ components of M3Y interaction potentials \cite{Sa79} supplemented by zero range potentials are given by the following : 

\begin{equation}
 t_{00}^{M3Y}(s, \epsilon) = 7999\frac{\exp( - 4s)}{4s} - 2134\frac{\exp( - 2.5s)}{2.5s} - 276 (1 - \alpha\epsilon)\delta(s)
\label{seqn4}
\end{equation}   
\noindent
and

\begin{equation}
  t_{01}^{M3Y}(s, \epsilon) =  -4886\frac{\exp( - 4s)}{4s} + 1176\frac{\exp( - 2.5s)}{2.5s} + 228 (1 - \alpha\epsilon)\delta(s),
\label{seqn5}
\end{equation}   
\noindent
where the energy dependence parameter $\alpha$=0.005/MeV. The zero-range potentials of Eqs.(4,5) represent the single-nucleon exchange term. The density dependent part appearing in Eq.(3) \cite{Ba05} has been taken to be of a general form

\begin{equation}
 g(\rho, \epsilon) = C (1 - \beta(\epsilon)\rho^n) 
\label{seqn6}
\end{equation}   
\noindent
which takes care of the higher order exchange effects and the Pauli blocking effects. This density dependence changes sign at high densities which is of crucial importance in fulfilling the saturation condition as well as giving different incompressibility $K_0$ values for SNM in ground state with different values of $n$ for the nuclear equation of state (EOS) \cite{Ba05}. The value of the parameter $n=2/3$ was originally taken by Myers in the single folding calculation \cite{My73}. The isospin asymmetry $X$ can be conveniently defined as 

\begin{equation}
 X = \frac{\rho_n-\rho_p}{\rho_n+\rho_p},~~~~\rho = \rho_n+\rho_p,
\label{seqn7}
\end{equation}   
\noindent
where $\rho_n$, $\rho_p$ and $\rho$ are the neutron, proton and nucleonic densities respectively. The asymmetry parameter $X$ can have values between -1  to +1, corresponding to pure proton matter and pure neutron matter respectively, while for SNM it becomes zero. For a single neutron interacting with rest of nuclear matter with isospin asymmetry $X$, the interaction energy per unit volume at $s$ is given by the following :

\begin{eqnarray}
 \rho_n v_{nn}(s)+\rho_p v_{np}(s)=&&\rho_n [v_{00}(s)+v_{01}(s)]+\rho_p[v_{00}(s)-v_{01}(s)]  \nonumber \\
  =&&[v_{00}(s)+v_{01}(s)X] \rho,
\label{seqn8}
\end{eqnarray}   
\noindent
while in case of a single proton interacting with rest of nuclear matter with isospin asymmetry $X$, the interaction energy per unit volume at $s$ is given by the following :

\begin{eqnarray}
 \rho_n v_{pn}(s)+\rho_p v_{pp}(s)=&&\rho_n [v_{00}(s)-v_{01}(s)]+\rho_p[v_{00}(s)+v_{01}(s)] \nonumber \\
=&&[v_{00}(s)-v_{01}(s)X] \rho.
\label{seqn9}
\end{eqnarray}   
\noindent
Summing the contributions for protons and neutrons and integrating over the entire volume of the infinite nuclear matter and multiplying by the factor $\frac{1}{2}$ to ignore the double counting in the process, the potential energy per nucleon $\epsilon_{pot}$ can be obtained by dividing the total potential energy by the total number of nucleons, 

\begin{equation}
 \epsilon_{pot}=\frac{g(\rho, \epsilon) \rho J_v}{2},
\label{seqn10}
\end{equation}   
\noindent
where

\begin{equation}
 J_v = J_{v00} + X^2 J_{v01} = \int \int \int [t_{00}^{M3Y}(s, \epsilon)+t_{01}^{M3Y}(s, \epsilon) X^2] d^3s . 
\label{seqn11}
\end{equation}   
\noindent

      Assuming interacting Fermi gas of neutrons and protons, the kinetic energy per nucleon $\epsilon_{kin}$ turns out to be  

\begin{equation}
 \epsilon_{kin} = [\frac{3\hbar^2k_F^2}{10m}] F(X), ~~~~F(X) = [\frac{(1+X)^{5/3} + (1-X)^{5/3}}{2}],
\label{seqn12}
\end{equation}   
\noindent
where $m$ is the nucleonic mass equal to 938.91897 $MeV/c^2$ and $k_F$, which becomes equal to Fermi momentum in case of the SNM, is given by the following :

\begin{equation}
 k_F^3 = 1.5\pi^2\rho.
\label{seqn13}
\end{equation}                                                                                                                                           
\noindent     

      The two parameters of Eq.(6), $C$ and $\beta$, are determined by reproducing the saturation conditions. It is worthwhile to mention here that due to attractive character of the M3Y forces the saturation condition for cold nuclear matter is not fulfilled. However, the realistic description of nuclear matter properties can be obtained with this density dependent M3Y effective interaction. Therefore, the density dependence parameters have been obtained by reproducing the saturation energy per nucleon and the saturation nucleonic density of the cold SNM.   

      The energy per nucleon $\epsilon=\epsilon_{kin}+\epsilon_{pot}$ obtained for the cold SNM for which $X=0$ is given by the following :

\begin{equation}
 \epsilon = [\frac{3\hbar^2k_F^2}{10m}] + \frac{g(\rho, \epsilon) \rho J_{v00}}{2},
\label{seqn14}
\end{equation}   
\noindent
where $ J_{v00}(\epsilon) = \int \int \int t_{00}^{M3Y}(s, \epsilon) d^3s$ represents the volume integral of the isoscalar part of the M3Y interaction supplemented by the zero-range potential. The Eq.(14) can be rewritten with the help of Eq.(6) as 

\begin{equation}
 \epsilon(\rho) = [\frac{3\hbar^2k_F^2}{10m}] + [\frac{\rho J_{v00} C (1 - \beta\rho^n)}{2}]  
\label{seqn15}
\end{equation}
\noindent
and differentiated with respect to $\rho$ to yield equation  

\begin{equation}
 \frac{\partial\epsilon}{\partial\rho} = [\frac{\hbar^2k_F^2}{5m\rho}] + \frac{J_{v00} C}{2} [1 - (n+1)\beta\rho^n]. 
\label{seqn16}
\end{equation}
\noindent
The equilibrium density of the cold SNM is determined from the saturation condition $\frac{\partial\epsilon}{\partial\rho} = 0$. Then Eq.(15) and Eq.(16) with the saturation condition can be solved simultaneously for fixed values of the saturation energy per nucleon $\epsilon_0$ and the saturation density $\rho_{0}$ of the cold SNM to obtain the values of the density dependence parameters $\beta$ and C. Density dependence parameters $\beta$ and C, thus obtained, can be given by the following :  

\begin{equation}
 \beta = \frac{[(1-p)\rho_{0}^{-n}]}{[(3n+1)-(n+1)p]},
\label{seqn17}
\end{equation} 
\noindent
where

\begin{equation}
 p = \frac{[10m\epsilon_0]}{[\hbar^2k_{F_0}^2]},
\label{seqn18}
\end{equation} 
\noindent
and 

\begin{equation}
 k_{F_0} = [1.5\pi^2\rho_0]^{1/3},
\label{seqn19}
\end{equation} 
\noindent

\begin{equation}
 C = -\frac{[2\hbar^2k_{F_0}^2] }{ 5mJ_{v00} \rho_0[1 - (n+1)\beta\rho_0^n]},
\label{seqn20}
\end{equation} 
\noindent
respectively. It is quite obvious that the density dependence parameter $\beta$ obtained by this method depends only on the saturation energy per nucleon $\epsilon_0$, the saturation density $\rho_{0}$ and the index $n$ of the density dependent part but not on the parameters of the M3Y interaction while the other density dependence parameter $C$ depends on the parameters of the M3Y interaction also through the volume integral $J_{v00}$. 

      The incompressibility $K_0$ of the cold SNM which is defined as   
  
\begin{equation}
 K_0 = k_F^2\frac{\partial^2\epsilon}{\partial{k_F^2}} \mid_{k_F=k_{F_0}} = 9\rho^2\frac{\partial^2\epsilon}{\partial\rho^2} \mid_{\rho=\rho_0}
\label{seqn21}
\end{equation}
\noindent
can be theoretically obtained using Eq.(13), Eq.(16) and Eq.(21) as

\begin{equation}
 K_{0} = [-(\frac{3\hbar^2k_{F_0}^2}{5m})] - [\frac{9 J_{v00} C n(n+1) \beta\rho_0^{n+1}}{2}].
\label{seqn22}
\end{equation} 
\noindent
Since the product $J_{v00} C$ appears in the above equation, a cursory glance reveals that the incompressibility $K_0$ depends only upon the saturation energy per nucleon $\epsilon_0$, the saturation density $\rho_{0}$ and the index $n$ of the density dependent part of the interaction but not on the parameters of the M3Y interaction.     
 
      The energy per nucleon for nuclear matter with isospin asymmetry $X$ can be rewritten as    

\begin{eqnarray}
 \epsilon(\rho,X) =&& [\frac{3\hbar^2k_F^2}{10m}] F(X) + (\frac{\rho J_v C}{2}) (1 - \beta\rho^n) \nonumber \\ 
 =&& [\frac{3\hbar^2k_F^2}{10m}] F(X) - [\frac{\rho}{\rho_{0}}] [\frac{J_v}{J_{v00}}] [\frac{\hbar^2k_{F_0}^2 (1 - \beta\rho^n)}{5m[1 - (n+1)\beta\rho_{0}^n]}],
\label{seqn23}
\end{eqnarray}
\noindent
where $J_v=J_{v00} + X^2 J_{v01}$ and $J_{v01}(\epsilon) = \int \int \int t_{01}^{M3Y}(s, \epsilon) d^3s$ represents the volume integral of the isovector part of the M3Y interaction supplemented by the zero-range potential.    

\section{Density dependence of nuclear symmetry energy and constraints from finite nuclei}
\label{section3}

      The nuclear symmetry energy (NSE) $E_{sym}(\rho)$ which is the energy required per nucleon to change the SNM to pure neutron matter (PNM) is defined as

\begin{equation}
 E_{sym}(\rho)=\epsilon(\rho,1) -\epsilon(\rho,0).
\label{seqn24}
\end{equation}
\noindent

      The calculations have been performed using the values of the saturation density $\rho_0=0.1533 fm^{-3}$ \cite{BSD90} and the saturation energy per nucleon $\epsilon_0=-15.26 MeV$ \cite{CSB05} for the SNM obtained from the co-efficient of the volume term of Bethe-Weizs\"acker mass formula which is evaluated by fitting the recent experimental and estimated atomic mass excesses from Audi-Wapstra-Thibault atomic mass table \cite{Au03} by minimizing the mean square deviation. For a fixed value of $\beta$, the parameters $\alpha$ and $C$ can have any possible simultaneous values as determined from SNM. Using the usual values of $\alpha=0.005 MeV^{-1}$ for the parameter of energy dependence of the zero range potential and $n=2/3$ \cite{Ba05}, the values obtained for the density dependence parameters $C$ and $\beta$ and the SNM incompressibility $K_0$ are 2.07, 1.668$fm^2$ and 293.4 MeV respectively. The general theoretical observation by Colo' et al. is that the non-relativistic \cite{Co04} and the relativistic \cite{CG04} mean field models predict for the $K_0$ values which are significantly different from one another, namely $\approx$ 220-235 MeV and $\approx$ 250-270 MeV respectively. Considering the uncertainties in the extractions of $\epsilon_0$ \cite{CSB05} and $\rho_{0}$ values from the experimental masses and electron scattering, present non-relativistic mean field model estimate for $K_0$ is rather close to the relativistic mean field model estimates. However if one uses $n=1/3$ instead of $n=2/3$ then the value of $K_0$ comes out to be 226.1 MeV which is close to the lower limit of the other non-relativistic calculations. Using Eq.(15) and Eq.(23), the NSE is given by

\begin{figure}[htbp]
\eject\centerline{\epsfig{file=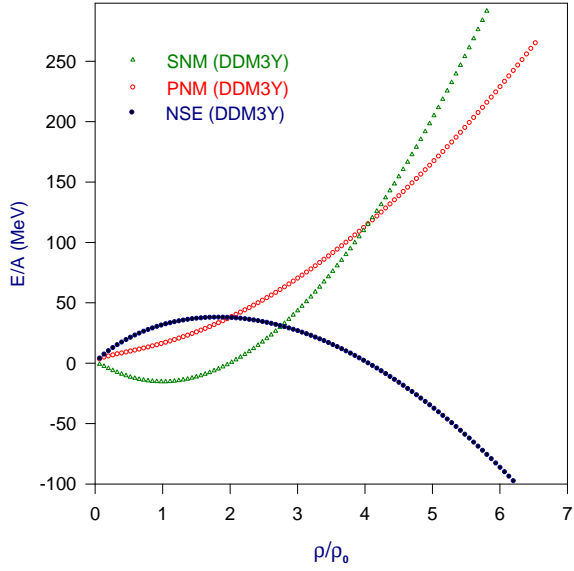,height=7.5cm,width=7.5cm}}
\caption
{The energy per nucleon $\epsilon$ = E/A of SNM (spin and isospin symmetric nuclear matter), PNM (pure neutron matter), NSE (nuclear symmetry energy $E_{sym}$) are plotted as functions of $\rho/\rho_0$ for the present calculations using DDM3Y interaction and the saturation energy per nucleon equal to -15.26 MeV.}
\label{fig1}
\end{figure}

\begin{equation}
 E_{sym}(\rho)= (2^{2/3} - 1)\frac{3}{5}E^0_F(\frac{\rho}{\rho_0})^{2/3}+\frac{C}{2} \rho (1 - \beta\rho^n) J_{v01},
\label{seqn25}
\end{equation}
\noindent
where the Fermi energy $E^0_F=\frac{\hbar^2k_{F_0}^2}{2m}$ for the SNM at ground state. The first term of the right hand side is the kinetic energy contribution with density dependence of $\rho^{2/3}$ whereas the second term arising due to nuclear interaction has a density dependence of the form of $a_1\rho+a_2\rho^{n+1}$ with $a_1$ and $a_2$ as constants with respect to density and values of $n$ are limited to 1/3 to 2/3 for any reasonable values of incompressibility. It is interesting to note that the effective interaction which provides unified description for radioactivity, nuclear matter and nuclear scattering with $n=2/3$ predicts the density dependence for the NSE to be of the form $a_0\rho^{2/3}+a_1\rho+a_2\rho^{5/3}$.

\begin{figure}[htbp]
\eject\centerline{\epsfig{file=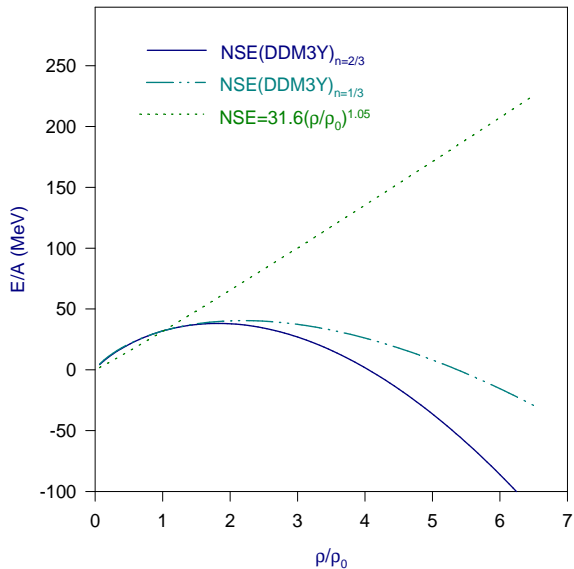,height=7.5cm,width=7.5cm}}
\caption
{The NSE (nuclear symmetry energy $E_{sym}$) as a function of $\rho/\rho_0$ for the present calculations using DDM3Y interaction is compared with the low energy behaviour extracted from isospin diffusion in heavy-ion collisions.}
\label{fig2}
\end{figure}

      In Fig.-1 plots of E/A for SNM, PNM and NSE as functions of $\frac{\rho}{\rho_0}$ are shown for $n=2/3$. The density dependence of the NSE at subnormal density from isospin diffusion \cite{Ch05} in heavy-ion collisions at intermediate energies has an approximate form of $31.6[\frac{\rho}{\rho_0}]^{1.05}$ MeV. In Fig.-2 this low energy behaviour of NSE $\approx 31.6[\frac{\rho}{\rho_0}]^{1.05}$ MeV is plotted along with NSE obtained using Eq.(25) for two extreme values of $n=1/3$ and $2/3$. At a density around $\rho\approx\rho_0$ both curves gives the same value of the NSE and at subnormal densities both forms are very close. At higher energies the present NSE using DDM3Y interaction peaks at $\rho\approx1.8\rho_0$ and becomes negative at $\rho\approx4\rho_0$. A negative NSE at high densities implies that the pure neutron matter becomes the most stable state. Consequently, pure neutron matter exists near the core of the neutron stars and since E/A for PNM is always positive, it is unbound by the nuclear force but bound due to gravitational attraction.  

\begin{figure}[htbp]
\eject\centerline{\epsfig{file=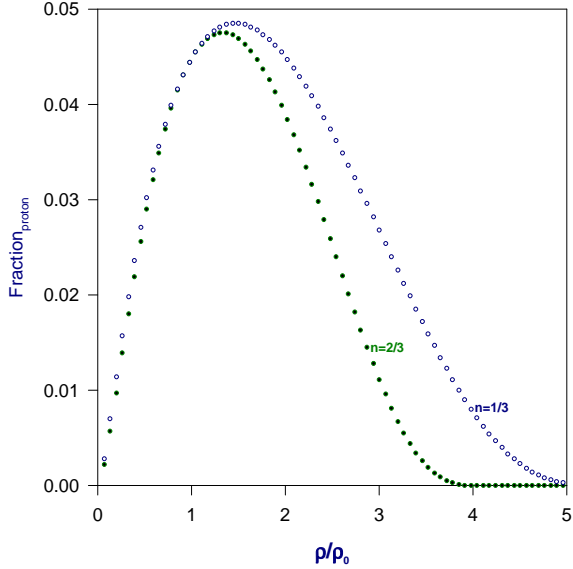,height=7.5cm,width=7.5cm}}
\caption
{The $\beta$ equilibrium proton fraction calculated with NSE (nuclear symmetry energy) obtained using DDM3Y interaction is plotted as a function of $\rho/\rho_0$.}
\label{fig3}
\end{figure}

\section{Cooling of neutron stars and the nuclear symmetry energy}
\label{section4}

      The knowledge of the density dependence of nuclear symmetry energy is important for understanding not only the stucture of radioactive nuclei but also many important issues in nuclear astrophysics, such as nucleosynthesis during presupernova evolution of massive stars and the cooling of protoneutron stars. A neutron star without neutrino trappings can be considered as a $n,p,e$ matter consisting of neutrons (n), protons (p) and electrons (e). The neutrinos do not accumulate in neutron stars and the $\beta$ equilibrium proton fraction $x_\beta$ [$=\rho_p/\rho$] is determined by \cite{La91} 

\begin{equation}
 \hbar c (3 \pi^2\rho x_\beta)^{1/3}= 4E_{sym}(\rho) (1 - 2 x_\beta). 
\label{seqn26}
\end{equation}
\noindent

The $\beta$ equilibrium proton fraction is therefore entirely determined by the NSE. The $\beta$ equilibrium proton fraction calculated using the present NSE is plotted as function of $\frac{\rho}{\rho_0}$ in Fig.-3. The maximum of $x_\beta\approx0.048 $ occurs at $\rho\approx1.4\rho_0$ and goes to zero at $\rho \approx 3.9\rho_0$ for $n=2/3$. The NSE extracted from the isospin diffusion in the intermediate energy heavy-ion collisions, having the approximate form of $31.6[\frac{\rho}{\rho_0}]^{1.05}$ MeV, provides a monotonically increasing $\beta$ equilibrium proton fraction and therefore can not be extended beyond normal nuclear matter densities. Present calculation, using NSE given by Eq.(25), of the $\beta$ equilibrium proton fraction forbids the direct URCA process since the equilibrium proton fraction is always less than 1/9 \cite{La91} which is consistent with the fact that there are no strong indications that fast cooling occurs. Although SNM incompressibilty is slightly on the higher side, yet the present calculations provide a rather `soft' nuclear symmetry energy. Using $n=1/3$ for which SNM incompressibility is 226.1 MeV and close to the lowest acceptable limit, results are very much similar for the $\beta$ equilibrium proton fraction excepting that it goes to zero at $\rho \approx 5.1\rho_0$.   

\section{Summary and conclusion}
\label{section5}

      In summary, we have calculated the symmetry energy using the isoscalar and the isovector components of M3Y effective NN interaction. The low density behaviour of the symmetry energy is found to be consistent with the nuclear symmetry energy extracted from the isospin diffusion in heavy-ion collisions at intermediate energies. Although the nuclear incompressibility is on the higher side, yet the present calculations provide a `soft' nuclear symmetry energy. The calculated $\beta$ equilibrium proton fraction forbids the direct URCA process which is consistent with the fact that there are no strong indications that fast cooling occurs. Present theoretical calculations provide density dependence of the nuclear symmetry energy that can be verified by the experimental signatures in several sensitive probes suh as the $\pi^-$ to $\pi^+$ ratio, tranverse collective flow and its excitation function as well as the neutron-proton differential flow.  

%\begin{references}

%\end{references}

\end{document}